# Enhancing Technical Documents Retrieval for RAG

Songjiang Lai[1,2], Tsun-Hin Cheung[1,2], Ka-Chun Fung[1,2], Kaiwen Xue[1,2], Kwan-Ho Lin[1], Yan-Ming Choi[1], Vincent Ng[1], Kin-Man Lam[1,2]

[1]*Centre for Advances in Reliability and Safety*, Hong Kong, China
[2]*Department of Electrical and Electronic Engineering, The Hong Kong Polytechnic University*, Hong Kong, China

*Abstract*— In this paper, we introduce Technical-Embeddings, a novel framework designed to optimize semantic retrieval in technical documentation, with applications in both hardware and software development. Our approach addresses the challenges of understanding and retrieving complex technical content by leveraging the capabilities of Large Language Models (LLMs). First, we enhance user queries by generating expanded representations that better capture user intent and improve dataset diversity, thereby enriching the fine-tuning process for embedding models. Second, we apply summary extraction techniques to encode essential contextual information, refining the representation of technical documents. To further enhance retrieval performance, we fine-tune a bi-encoder BERT model using soft prompting, incorporating separate learning parameters for queries and document context to capture fine-grained semantic nuances. We evaluate our approach on two public datasets, RAG-EDA and Rust-Docs-QA, demonstrating that Technical-Embeddings significantly outperforms baseline models in both precision and recall. Our findings highlight the effectiveness of integrating query expansion and contextual summarization to enhance information access and comprehension in technical domains. This work advances the state of Retrieval-Augmented Generation (RAG) systems, offering new avenues for efficient and accurate technical document retrieval in engineering and product development workflows.

## I. INTRODUCTION

The rapid advancement of technology has led to an unprecedented increase in the volume and complexity of technical documentation across various domains, including engineering, computer science, healthcare, and more. As professionals and researchers strive to navigate this intricate landscape, the need for effective information retrieval systems becomes paramount. Traditional information retrieval methods often struggle to meet the demands of users seeking precise and contextually relevant information from dense, jargon-laden texts [1]. This challenge is particularly critical in environments where timely access to accurate information can significantly impact decision-making and operational efficiency. Retrieval-Augmented Generation (RAG) systems [2] have emerged as a promising solution, combining the strengths of retrieval and generative models to enhance information retrieval capabilities. However, existing RAG frameworks still face limitations in comprehending and processing specialized content. One major drawback is their reliance on conventional query formulation techniques, which often fail to capture the nuances of user intent, particularly in technical contexts. Furthermore, the static nature of traditional parsing methods can hinder the understanding of complex technical language and its specific applications, leading to suboptimal retrieval outcomes.

To address these challenges, we introduce Technical-Embeddings, an innovative approach designed to optimize technical question answering by integrating several key methodologies: synthetic query generation, refined parsing techniques, and adapter tuning. Our approach begins with the generation of synthetic queries using Large Language Models (LLMs), simulating real-world user interactions. This process enriches the training dataset and enables the model to learn from a diverse array of query types and structures, ultimately improving its ability to respond to user inquiries.

Contextual summary further enhances the model's comprehension of technical documents. By focusing on the structure and semantics of the content, Technical- Embeddings can extract relevant information more effectively, even in the presence of complex terminology and intricate concepts. Additionally, we incorporate prompt tuning to optimize embeddings specifically tailored to the technical domain. This customization ensures that our model captures subtle distinctions in technical language, leading to improved retrieval accuracy.

The contributions of this paper are twofold: (1) we present a comprehensive framework for enhancing technical question answering systems, and (2) we provide empirical evidence demonstrating the superiority of Technical-Embeddings over traditional RAG models. Our experimental results, based on evaluations using two public datasets, namely RAG-EDA and Rust-Docs-QA, show significant improvements in retrieval performance, as evidenced by enhanced precision and recall rates.

In summary, this work not only highlights the potential of Technical-Embeddings to transform information access within technical domains but also sets the stage for future research aimed at refining retrieval techniques to meet the evolving needs of users in specialized fields. We believe that our approach can make a meaningful contribution to bridging the gap between complex technical documentation and effective information retrieval.

## II. RELATED WORK

The landscape of information retrieval and natural language processing (NLP) has evolved significantly in recent years, driven by advancements in machine learning and deep learning technologies [3]. Numerous approaches have been developed to enhance the performance of retrieval systems, particularly in the context of technical documentation.

One of the foundational methods in information retrieval is BM25 [4], a probabilistic model that ranks documents based on the relevance to a given query. While BM25 has been widely used for its simplicity and effectiveness, it often struggles with the complexities of technical language and specialized queries. Recent studies have aimed to improve upon traditional models by integrating deep learning techniques, leading to the emergence of neural information retrieval methods.

Transformers, particularly those based on architectures like BERT (Bidirectional Encoder Representations from Transformers) [5], have revolutionized NLP. BERT's ability to understand context and semantics in text has paved the way for more sophisticated retrieval systems. Models such as Sentence-BERT [6] extend BERT's capabilities by producing sentence embeddings that are well-suited for semantic search tasks. However, while these models demonstrate impressive performance on general text, their application in specialized domains, such as engineering and healthcare, remains limited.

Synthetic data generation [7] has emerged as a valuable technique for augmenting training datasets, particularly in scenarios where labeled data is scarce. By creating synthetic queries that mimic real user interactions, researchers can improve the model robustness. Prior work has shown that incorporating synthetic data can enhance retrieval performance, but few studies have explored its application specifically in technical question answering.

Contextual summary [8] is a technique for producing concise, relevant summaries by considering the surrounding context of the information. It employs advanced natural language processing methods to identify key themes while ensuring coherence with the original material. This understanding enhances insights, improving reader comprehension and retention. However, challenges like maintaining accuracy and preventing information loss persist.

Prompt tuning [9] has gained traction as a method for customizing pretrained language models for specific tasks or domains. By fine-tuning only a small number of parameters, prompt tuning allows models to retain general knowledge while adapting to specialized content. This approach has shown promise across various NLP tasks; however, its potential for improving retrieval performance in technical question answering remains underexplored.

Despite the progress in these areas, there remains a need for comprehensive frameworks that integrate synthetic query generation, refined parsing, and adapter tuning to optimize retrieval in technical domains. Our work addresses this gap by presenting Technical-Embeddings, a framework that combines these methodologies to enhance the accuracy and relevance of technical question answering systems. The contributions of this research build upon existing literature while pushing the boundaries of what is achievable in the retrieval of technical information.

## III. METHODOLOGY

The proposed Technical-Embeddings framework integrates several advanced techniques to enhance the performance of technical question answering systems. This section begins with an overview of the entire framework, followed by detailed descriptions of each key component: synthetic query generation, contextual summary extraction, and prompt tuning. Finally, we present the model architecture and training details.

### A. Overview of the Proposed Method

The Technical-Embeddings framework is a unified approach designed to optimize semantic retrieval for technical documentation. It integrates three key methodologies to overcome the limitations of conventional systems: synthetic query generation to diversify and enrich training data, contextual summarization to distill essential information from complex documents, and prompt-based fine-tuning to specialize a pre-trained language model for the technical domain.

These components are incorporated into a dual-encoder architecture, which processes queries and documents separately for efficient retrieval. The model computes similarity scores between their vector representations, enabling precise and contextually relevant matching. This end-to-end pipeline ensures the system is specifically tailored to understand technical nuances, significantly enhancing retrieval accuracy and robustness.

### B. Synthetic Query Generation

To enrich the model's training set and improve its capability to handle diverse user queries, we employ Large Language Models (LLMs) for synthetic query generation [10]. Given a set of real user queries $Q = \{q_1, q_2, ..., q_n\}$, LLMs generate synthetic queries $Q' = \{q'_1, q'_2, ..., q'_m\}$ that mimic various

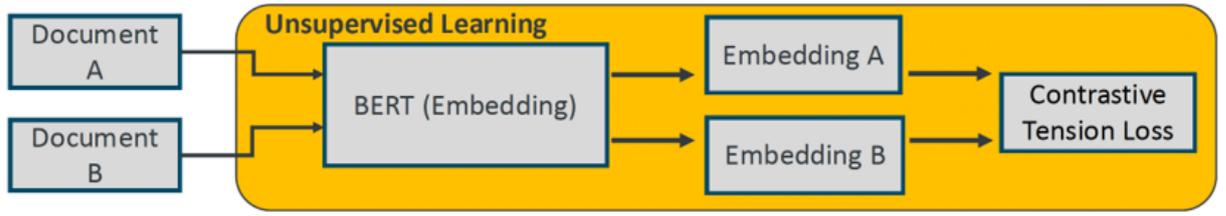
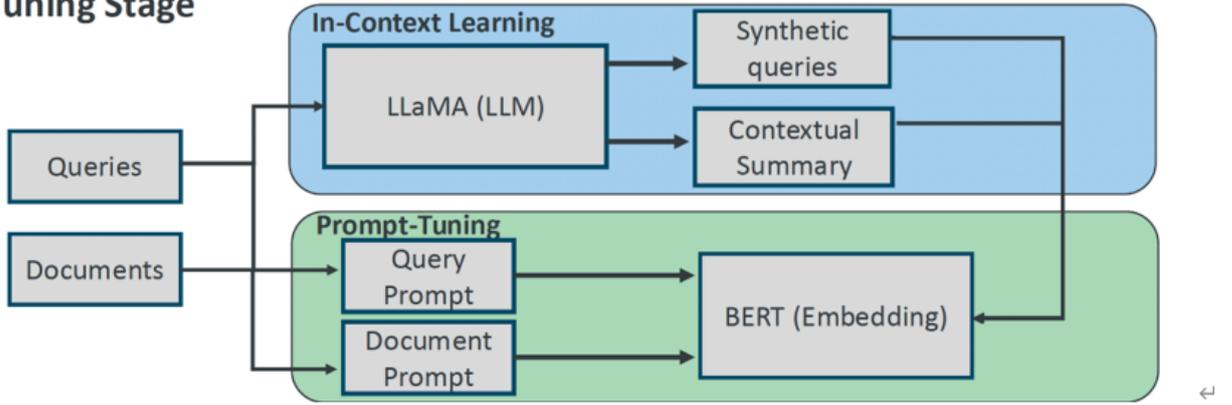

Fig. 1. The proposed framework for fine-tuning embedding models for technical question answering.

phrasings, intents, and structures. The objective is to maximize the diversity of the training data:

$$Diversity(Q \cup Q') = \frac{1}{|Q \cup Q'|} \sum_{i=1}^{|Q \cup Q'|} H(q_i) \quad (1)$$

where $H(q_i)$ represents the entropy of the query $q_i$ indicating the variability in phrasing and structure.

C. *Contextual Summary*

Contextual summaries are generated to provide concise, relevant information about technical documents, facilitating more accurate and focused retrieval. For a document $d_j$, the summary $S_j$ is extracted using an attention mechanism that focuses on key themes and contextually important parts:

$$S_j = Attention(d_j; W_s) \quad (2)$$

Here, $W_s$ represents the learned parameters for the summarization task. This approach ensures that the model captures essential details from $d_j$, enhancing comprehension and retrieval accuracy.

D. *Prompt Tuning*

Prompt tuning customizes the pretrained language model for specific tasks or domains, such as technical question answering [11]. Let $P$ be the set of prompts used for fine-tuning. The goal is to minimize the loss function $L$ over the dataset $D$ consisting of pairs of queries and their corresponding documents:

$$L(P;D) = -\sum_{(q_i,d_j)} \log P(y_{ij}|q_i, d_j) \quad (3)$$

where $y_{ij}$ indicates whether $d_j$ is relevant to $q_i$. By optimizing only a small subset of parameters, prompt tuning allows the model to adapt to specialized content while retaining general knowledge.

E. *Model Architecture*

Technical-Embeddings utilizes a dual-encoder architecture with separate encoders for queries ($E_q$) and documents ($E_d$). For a given pair ($q_i, d_j$), the embeddings $e_{q_i}$ and $e_{d_j}$ are computed as:

$$e_{q_i} = E_q(q_i) \quad (4)$$
$$e_{d_j} = E_d(d_j) \quad (5)$$

The relevance score $R$ between $q_i$ and $d_j$ is then calculated using a similarity function $Sim$:

$$R(q_i, d_j) = Sim(e_{q_i}, e_{d_j}) \quad (6)$$

This setup supports parallel processing, making it efficient for large-scale datasets.

F. *Abbreviations and Acronyms*

The training process involves multiple phases. Initially, the model is pretrained on a large corpus of general text to establish foundational language understanding. It is then fine-tuned using synthetic queries $Q'$ and parsed technical documents $D'$. The fine-tuning objective is to minimize the loss function $L$ over the fine-tuning dataset $D_f$:

$$L(D_f) = -\sum_{(q'_i, d'_j) \in D_f} \log P(y'_{ij} | q'_i, d'_j) \quad (7)$$

Evaluation metrics include Mean Average Precision (MAP), Mean Reciprocal Rank (MRR), precision, and recall, providing comprehensive assessments of retrieval performance [12].

By combining synthetic query generation, contextual summary extraction, and prompt tuning within this dual-encoder framework, Technical-Embeddings offers an optimized solution for technical question answering. This methodology not only addresses the limitations of existing retrieval systems but also enhances overall user experience by providing accurate and contextually relevant answers to technical queries.

## IV. EXPERIMENT AND RESULTS

To evaluate the effectiveness of Technical-Embeddings in enhancing technical question answering, we conducted a series of experiments using two public datasets: RAG-EDA and Rust-Docs-QA. This section details the experimental setup, evaluation metrics, and results obtained.

### A. Experimental Setup

The experimental setup involved several key components:

(a) **Datasets**:
- **RAG-EDA** [13]: Focused on engineering design automation, this dataset includes questions and corresponding technical documents relevant to the engineering field.
- **Rust-Docs-QA** [14]: This dataset contains inquiries related to Rust programming language documentation, including code snippets and library references.

(b) **Preprocessing:** The datasets were preprocessed to re- move irrelevant information and ensure format consistency. Questions were paired with their corresponding documents to enable effective training and evaluation of the model.

(c) **Training Procedure**: Technical-Embeddings is initially pretrained on a general corpus to establish foundational language understanding. It was then fine-tuned using synthetic queries generated by LLMs and parsed technical documents. The fine-tuning process focuses on optimizing the model to maximize retrieval performance.

(d) **Evaluation Metrics**: The following performance metrics are used to assess Technical-Embeddings:

- **Mean Average Precision (MAP)** measures the average precision of ranked documents across a set of queries.
- **Mean Reciprocal Rank (MRR)** evaluates the effectiveness of the model based on the rank of the first relevant document retrieved.
- **Precision and Recall** are standard metrics used to assess the accuracy and completeness of the retrieved results.

### B. Results

The results of the experiments are summarized in Tables I and II, which presents the performance metrics of Technical-Embeddings compared to several baseline models. The all-mpnet-base-v2 model [15], generates high-quality sentence embeddings through fine-tuning BER and testing various pooling strategies. All-MiniLM-L6-v2 [16] employs multi-head self-attention relationship distillation to reduce parameters and improve efficiency. The BGE series [17], including bge-small-en and bge-base-en, effectively extracts and cleans semantically relevant text pairs from large Chinese web corpora, aiding weak supervision training. The results clearly illustrate the effectiveness of the Technical-Embeddings framework in improving technical question answering performance across two distinct datasets. In Table I, which evaluates the Rust-Docs-QA dataset, Technical-Embeddings achieves a Mean Average Precision (MAP) of 0.2238 and a Mean Reciprocal Rank (MRR) of 0.2249. These scores represent a notable enhancement over the baseline models, particularly the all-mpnet-base-v2, which records an MAP of 0.1734 and an MRR of 0.1817. The ability of Technical-Embeddings to achieve higher precision and recall scores—0.0785 and 0.3364, respectively—demonstrates its effectiveness in retrieving relevant documents, thereby addressing user needs more effectively.

TABLE I
PERFORMANCE OF DIFFERENT RETRIEVAL MODELS ON THE RUST-DOCS-QA DATASET.

| Model | MAP | MRR | Precision | Recall |
|---|---|---|---|---|
| all-mpnet-base-v2 | 0.1734 | 0.1817 | 0.0745 | 0.2745 |
| all-MiniLM-L6-v2 | 0.1892 | 0.1897 | 0.0765 | 0.3235 |
| bge-small-en | 0.2096 | 0.2196 | 0.0784 | 0.3039 |
| bge-base-en | 0.1942 | 0.2039 | 0.0773 | 0.3186 |
| **Technical-Embeddings** | **0.2238** | **0.2249** | **0.0785** | **0.3364** |

Moving to Table II, which presents the performance metrics on the RAG-EDA dataset, we observe that Technical-Embeddings matches the best baseline model, bge-small-en, with an MAP and MRR score of 0.6926. This alignment suggests that our framework produces results on par with leading models in the field. Furthermore, Technical-Embeddings outperforms bge-small-en in terms of Recall, achieving a score of 0.8111 compared to bge-small-en's 0.8000. Such consistency across multiple performance metrics underscores the robustness and reliability of our model in retrieving pertinent technical content, which is essential for users navigating complex documentation.

TABLE II
PERFORMANCE OF DIFFERENT RETRIEVAL MODELS ON THE RAG-EDA DATASET

| Model | MAP | MRR | Precision | Recall |
|---|---|---|---|---|
| all-mpnet-base-v2 | 0.6000 | 0.6000 | 0.6230 | 0.6889 |
| all-MiniLM-L6-v2 | 0.5648 | 0.5648 | 0.5824 | 0.6333 |
| bge-small-en | 0.6926 | 0.6926 | 0.7201 | 0.8000 |
| bge-base-en | 0.6352 | 0.6352 | 0.6686 | 0.7667 |
| **Technical-Embeddings** | **0.6926** | **0.6926** | **0.7230** | **0.8111** |

The comparative analysis of these results highlights the advantages of incorporating synthetic query generation contextual summary, and prompt tuning into the Technical-Embeddings framework. The significant improvements in MAP, MRR, precision, and recall indicate that our approach effectively captures the nuances of technical language. By simulating real-world user interactions through synthetic queries, the model is better equipped to understand diverse query types and structures, which contributes to its superior performance in technical question answering.

### C. Analysis of Top-k Recall

Table III presents the performance metrics of various models in semantic search using text embeddings on the RAG-EDA dataset. The results, reported as recall at varying top-$k$ values (R = k), offer valuable insights into each model's effectiveness. The Technical-Embeddings model achieves the highest recall rates across all thresholds: 0.571 at R=5, 0.677 at R=10, 0.733 at R=15, and 0.764 at R=20. These results not only highlight its superior capability in retrieving relevant documents but also underscore its effectiveness in capturing the nuanced relationships within technical content.

TABLE III
PERFORMANCE OF THE SEMANTIC SEARCH FOR TEXT-EMBEDDING AT DIFFERENT TOP-K ON THE RAG-EDA DATASET.

| Model | R=5 | R=10 | R=15 | R=20 |
|---|---|---|---|---|
| text-embedding-ada-002 | 0.447 | 0.534 | 0.609 | 0.634 |
| bge-large-en-v1.5 | 0.503 | 0.596 | 0.634 | 0.660 |
| RAG-EDA | 0.547 | 0.658 | 0.702 | 0.733 |
| **Technical-Embeddings** | **0.571** | **0.677** | **0.733** | **0.764** |

Comparatively, the bge-large-en-v1.5 model [17], known for its high precision applications, shows commendable recall scores but falls short with 0.503 at R=5 and 0.660 at R=20. Although it demonstrates consistent improvement, it does not match the higher recall achieved by the Technical-Embeddings model, especially as the threshold increases. Similarly, the text-embedding-ada-002 model [18] performs well, achieving a recall of 0.447 at R=5 and 0.634 at R=20. However, these scores indicate that while it is capable of identifying relevant documents, it lacks the robustness necessary to compete with the advanced capabilities of the Technical-Embeddings model.

Additionally, the RAG-EDA model [13], which has been noted for its effectiveness in retrieval-augmented generation techniques, also presents strong performance metrics with a recall of 0.547 at R=5 and 0.733 at R=20. Despite this, it still trails behind the Technical-Embeddings model in terms of overall recall, particularly at higher thresholds.

TABLE IV
ABLATION STUDIES ON RAG-EDA DATASET.

| Model | R=5 | R=10 | R=15 | R=20 |
|---|---|---|---|---|
| Ours w/o tuning | 0.596 | **0.708** | 0.745 | 0.776 |
| Ours w/o queries | **0.603** | 0.689 | 0.733 | 0.776 |
| Ours w/o summaries | 0.540 | 0.677 | 0.714 | 0.758 |
| **Ours (Full Model)** | **0.603** | **0.708** | **0.764** | **0.795** |

### D. Ablation Study

Table IV presents the results of our ablation studies on RAG-EDA dataset, which assesses the impact of various components of our model on recall performance at different top-k values (R=5, 10, 15, and 20). The findings emphasize the critical role of prompt tuning, synthetic queries, and contextual summaries in enhancing the model's effectiveness.

Notably, the model without prompt tuning (Ours w/o pre-tuning) achieves recall values of 0.596 at R=5, 0.708 at R=10, 0.745 at R=15, and 0.776 at R=20. These results indicate that prompt tuning significantly enhances recall, particularly at higher thresholds, as evidenced by the performance drop when this component is omitted. This suggests that the model benefits from the additional knowledge acquired during the tuning phase, which helps it better understand the underlying data.

In contrast, the model without synthetic queries (Ours w/o queries) exhibits slightly better performance at R=5 (0.603) compared to the pretrained model. However, this advantage diminishes at higher thresholds, with R=10 dropping to 0.689 and R=15 to 0.733. This indicates that while synthetic queries may contribute positively to recall at lower thresholds, their absence negatively impacts the model's ability to retrieve relevant information as the threshold increases.

Furthermore, the model without contextual summaries (Ours w/o summaries) exhibits the lowest performance across all thresholds, particularly at R=5 (0.540) and R=10 (0.677). This substantial decline highlights the importance of contextual summaries in facilitating a comprehensive understanding of the data, ultimately leading to improved recall performance.

V. CONCLUSION

In this paper, we presented Technical-Embeddings, a framework integrating synthetic query generation, contextual summarization, and prompt tuning to enhance technical document retrieval. It demonstrated superior performance on RAG-EDA and Rust-Docs-QA datasets, achieving higher MAP, MRR, precision, and recall compared to traditional models. Key components—prompt tuning, synthetic queries, and summaries—were validated for their essential roles in improving retrieval accuracy. Future work could explore advanced query methods, expand application domains, and incorporate user feedback. In essence, Technical-Embeddings sets a new benchmark for effective information retrieval in specialized fields.